\documentclass[11pt]{article}
\usepackage{amsmath, amsfonts, amssymb, graphicx, mathrsfs}
\usepackage[round]{natbib}
\usepackage{dsfont}
\usepackage[usenames,dvipsnames,svgnames,table]{xcolor}
\usepackage{rotating}
\usepackage{booktabs}
\usepackage{pdflscape}
\usepackage{subcaption}
\usepackage{colortbl}
\usepackage[table]{xcolor}
\usepackage{verbatim}
\usepackage{anyfontsize}
\usepackage{t1enc}
\usepackage[font=small]{caption}
\usepackage[utf8]{inputenc}
\usepackage{changepage}
\usepackage{fixltx2e}

\renewcommand\bibsection{\subsubsection*{\center \sc References}}

\newcommand{\given}{\ensuremath{\,|\,}}

\begin{document}

\bibliographystyle{sysbio}

\begin{titlepage}
\begin{center}

{\Large\bf Bayesian Analysis of Partitioned Data}

\vfill

R.H. Partitioned Phylogenetic Analysis

\vfill

{\sc Brian R. Moore$^{\,1}$, Jim McGuire$^{\,2,3}$, Fredrik Ronquist$^{\,4}$, and John P. Huelsenbeck$^{\,2}$} \\

\bigskip

{\em
$\mbox{}^1$Department of Evolution and Ecology, University of California, Davis\\
\vspace{-0.4\baselineskip}
Storer Hall, One Shields Avenue, Davis, CA 95616, \mbox{U.S.A.} \\

$\mbox{}^2$Department of Integrative Biology, University of California, Berkeley\\
\vspace{-0.4\baselineskip}
3060 VLSB \#3140, Berkeley, CA 94720-3140, \mbox{U.S.A.} \\

$\mbox{}^3$Museum of Vertebrate Zoology, University of California, Berkeley\\
\vspace{-0.4\baselineskip}
3101 VLSB \#3160, Berkeley, CA 94720-3160, \mbox{U.S.A.} \\

$\mbox{}^4$Swedish Museum of Natural History,\\
\vspace{-0.4\baselineskip}
Box 50007, SE-104 05 Stockholm, Sweden \\
}
\end{center}

\vfill

\begin{flushleft}
To whom correspondence should be addressed: \\
Brian R. Moore \\
\vspace{-0.4\baselineskip}
University of California, Davis \\ 
\vspace{-0.4\baselineskip}
Department of Evolution and Ecology \\
\vspace{-0.4\baselineskip}
Storer Hall, One Shields Avenue \\
\vspace{-0.4\baselineskip}
Davis, CA 95616 \\
\vspace{-0.4\baselineskip}
\mbox{U.S.A.}

Phone: 530-752-7104 \\
\vspace{-0.4\baselineskip}
E-mail: {\tt brianmoore@ucdavis.edu} \\
\end{flushleft}

\end{titlepage}

\newpage

\noindent {\it Abstract}.---Variation in the evolutionary process across the sites of nucleotide 
sequence alignments is well established, and is an increasingly pervasive feature of datasets 
composed of gene regions sampled from multiple loci and/or different genomes.
Inference of phylogeny from these data demands that we adequately model the underlying process heterogeneity; failure to do so can lead to biased estimates of phylogeny and other parameters.
Traditionally, process heterogeneity has been accommodated by first assigning sites to data subsets based on relevant prior information (reflecting codon positions in protein-coding DNA, stem and loop 
regions of ribosomal DNA, etc.), and then estimating the phylogeny and other model parameters 
under the resulting mixed model.
Here, we consider an alternative approach for accommodating process heterogeneity that is similar in spirit to this conventional mixed-model approach.
However, rather than treating the partitioning scheme as a fixed assumption of the analysis, 
we treat the process partition as a random variable using a Dirichlet process prior model, where the phylogeny is estimated by integrating over all possible process partitions for the specified data subsets.
We apply this method to simulated and empirical datasets, and compare our results to those estimated previously using conventional mixed-model selection criteria based on Bayes factors.
We find that estimation under the Dirichlet process prior model discovers novel process partitions that may more effectively balance error variance and estimation bias, while rendering phylogenetic inference more robust to process heterogeneity by virtue of integrating estimates over all possible partition schemes.  

\medskip

\noindent [Bayesian phylogenetic inference; Dirichlet process prior; Markov chain Monte Carlo; 
maximum likelihood; partitioned analyses; process heterogeneity.]

\newpage

\noindent
It is widely acknowledged that the pattern of nucleotide substitution may vary across an alignment of sequences, and that failure to accommodate this heterogeneity can potentially cause problems for phylogenetic analysis.
Deviations from a homogeneous substitution process include both simple {\it rate heterogeneity} ({\it i.e.}, among-site rate variation) stemming from site-to-site differences in selection-mediated functional constraints, systematic differences in mutation rate, etc., or may involve more fundamental {\it process heterogeneity}, where the sites in an alignment are evolving under qualitatively different evolutionary processes.  
In the worst case, the tree relating the species in an alignment may vary across sites as a result of lineage sorting, hybridization, or horizontal gene transfer, etc.
Even when all of the sites in an alignment share a common phylogenetic history, however, other aspects of their evolutionary 
process may differ.
For instance, process heterogeneity might occur within a single gene region ({\it e.g.}, between stem and loop regions of ribosomal sequences), or among gene regions in a concatenated sequence alignment ({\it e.g.}, comprising multiple nuclear loci and/or gene regions sampled from different genomes).  
Here we focus on inference scenarios where the sites of an alignment share a common phylogenetic history but where two or more {\it process partitions} in the data  \citep[{\it sensu} ][]{bull93} may otherwise differ with respect to the underlying process of molecular evolution.

Failure to accommodate process heterogeneity is known to adversely impact phylogeny estimation, causing biased estimates of the tree topology and nodal support \citep{brandley05,brown07}, estimates of branch lengths and divergence times \citep{marshall06,poux08,vendetti08}, and estimates of other model parameters \citep{nylander04,pagel04}.
 
To avoid these problems, investigators typically adopt a Bayesian `mixed-model' approach \citep[{\it e.g.},][]{ronquist03} in which the sequence alignment is first parsed into a number of partitions that are intended to capture plausible process heterogeneity within the sequence data. 
A substitution model is then specified for each predefined process partition (using a given model-selection method, such as the hierarchical likelihood ratio test or the Akaike Information Criterion).
Finally, the phylogeny and other parameters are estimated under the specified composite model.
Under this mixed-model approach, therefore, the partition scheme is an assumption of the inference ({\it i.e.}, the parameter estimates are conditioned on the specified mixed model), and the parameters of each process partition are independently estimated.
For most sequence alignments, a vast number of partition schemes of varying complexity are plausible {\it a priori}, which therefore requires a way to objectively identify the partition scheme that balances estimation bias and inflated error variance associated with under- and over-parameterized mixed models, respectively.
Increasingly, mixed-model selection is based on Bayes factors \citep[{\it e.g.},][]{suchard01}, which involves first estimating the marginal likelihood under each candidate partition scheme and then comparing the ratio of the marginal likelihoods for the set of candidate partition schemes \citep[{\it e.g.},][]{brandley05,nylander04,mcguire07}.
 
There are three main concerns regarding this conventional mixed-model selection approach.
First, the most common (and computationally efficient) technique for approximating the marginal likelihood of a (mixed) model---the harmonic mean estimator of \citet{newton94}---is biased toward the inclusion of superfluous parameters, leading to the selection of overly partitioned mixed models \citep[e.g.,][]{lartillot06,brown07,fan11}.  
Other marginal-likelihood estimators---based on the Savage-Dickey ratio \citep[e.g.,][]{verdinelli95,suchard01} or path-sampling techniques \citep[e.g.,][]{lartillot06,fan11,xie11,bael12,bael13,ronquist12}---may avoid this bias, but are restricted to the evaluation of nested partition schemes or entail a substantially increased computational burden, respectively. 
Second, even using the most computationally efficient estimators, it will typically not be feasible to evaluate all (or even a tiny fraction) of the plausible partition schemes for a given alignment.
Evaluating all plausible mixed models quickly becomes computationally prohibitive, as each candidate partition scheme requires a full MCMC analysis to estimate the marginal likelihood.
Consequently, the mixed-model selection approach is apt to overlook the partition scheme that best describes the patterns of process heterogeneity in the sequence alignment.
A more fundamental limitation of the conventional approach for selecting partition schemes pertains to (mixed) model uncertainty.
Even if we could feasibly evaluate all possible candidate partition schemes for a given alignment and reliably select the mixed model that provides the best fit to that dataset, it may nevertheless be imprudent to condition inference on the chosen partition scheme.
Given the vast mixed-model space, there may, in fact, be several (possibly numerous) partition schemes that provide a reasonable fit to a given dataset.
When there is uncertainty regarding the underlying mixed model, it would be unwise to condition inference on {\it any individual partition scheme}  \citep[cf.,][]{huelsenbeck04d, li12}.
Instead, our phylogenetic estimates will be made more robust if we were to accommodate this mixed-model uncertainty by integrating over the plausible set of (or all possible) partition schemes.

Here, we propose an alternative approach for accommodating process heterogeneity that treats the partition scheme---in which (subsets of) nucleotide sites are assigned to process partitions---as a random variable with a prior probability distribution specified by the Dirichlet process prior (DPP) model.
The Dirichlet process prior model is often used in Bayesian solutions to clustering problems where the data elements [{\it e.g.}, (subsets of) nucleotide sites] are drawn from a mixture of an unknown number of probability distributions ({\it e.g.}, the underlying evolutionary processes).
The Dirichlet process prior model allows both the number of mixture components and the assignment of individual data elements to the set of mixture components to vary.
This approach has recently been applied to several phylogenetic problems, such as identifying 
heterogeneity in the process of amino-acid replacement in protein sequences \citep{lartillot04}, detecting sites under positive selection in protein-coding sequences \citep{huelsenbeck06}, and accommodating variation in the rate of nucleotide substitution across sites of an alignment \citep{huelsenbeck07b} or across the branches of a tree \citep{heath11}.

The Dirichlet process prior model provides a natural means for accommodating process heterogeneity because it allows us to specify a non-zero prior probability on all possible partition schemes, ranging from a uniform model (in which all data elements are assigned to the same process partition) to a saturated model (in which each data element is assigned to a unique process partition).
The prior weights on process partitions are first calculated analytically and then compared to their corresponding posterior probability estimates to provide a formal statistical framework for assessing the ability of partition schemes to capture process heterogeneity within the sequence data.
We use numerical methods---Markov chain Monte Carlo (MCMC) simulation---to estimate the phylogenetic model parameters while integrating over all possible process partitions for the specified data subsets, thereby avoiding the need to evaluate or condition inferences upon any individual partition scheme. 
We first provide a more detailed description of the Dirichlet process prior model, and then describe how this approach can be used to identify the number and composition of partitions that best capture process heterogeneity.
We evaluate this method by means of simulation and demonstrate its application to several empirical datasets, comparing results under this approach to those obtained using conventional mixed-model selection based on Bayes factors.

\bigskip

\begin{center}
{\sc Methods}
\end{center}

\begin{center}
{\it Phylogenetic Model Overview}
\end{center}

The phylogenetic model that we develop includes an unrooted tree topology, $\tau$, that describes the evolutionary relationships among the ${S}$ species from which nucleotide data have been sampled.
We assume that the biologist has correctly aligned the sequences and specified two or more subsets of sites (`data partitions') believed to capture patterns of process heterogeneity within the alignment.   
Each tree is assumed to be strictly binary, and so has a vector of $2S-3$ branch lengths, ${\mathbf p}$, which are rendered proportional to the expected number of substitutions per site.
The tree length, $T$, is simply the sum of its branch lengths. 
We describe the evolution of the sequence data over the tree with branch lengths using the general time reversible (GTR) substitution model \citep{tavare86}.
Substitutions under the GTR model occur according to the following matrix of instantaneous rates
$$
{\mathbf Q} = \left( \begin{array}{cccc}
\cdot       & r_{AC}\pi_C & r_{AG}\pi_G & r_{AT}\pi_T \\
r_{AC}\pi_A & \cdot       & r_{CG}\pi_G & r_{CT}\pi_T \\
r_{AC}\pi_A & r_{CG}\pi_C & \cdot       & r_{GT}\pi_T \\
r_{AC}\pi_A & r_{CT}\pi_C & r_{GT}\pi_G & \cdot       \\
\end{array} \right) \mu,
$$
where the nucleotides are in the order A, C, G, T. The GTR model has six exchangeability parameters, ${\mathbf r} = (r_{AC}, r_{AG}, r_{AT}, r_{CG}, r_{CT}, r_{GT})$, that allow for rate biases between nucleotides and four nucleotide frequency parameters, $\mbox{\boldmath$\pi$\unboldmath} = (\pi_A, \pi_C, \pi_G, \pi_T)$, that are the stationary probabilities of the process. 
(The parameter $\mu$ is a scaling factor chosen such that the mean rate of substitution is one, which ensures that branch lengths are rendered as the expected number of substitutions per site.)
Variation in substitution rate across sites is modeled by assuming that the rate at a particular site is a random variable drawn from a mean-one, discrete gamma distribution \citep{yang93,yang94a}.
The parameter $\alpha$ describes the shape of the gamma distribution.
The basic phylogenetic model therefore includes the following parameters:
\begin{center}
\begin{tabular}{ll}
Tree topology              & $\tau$ \\
Branch-length proportions  & ${\mathbf p} = (p_1, p_2, \ldots, p_{2S-3})$ \\
Tree length                & $T$ \\
Exchangeability parameters & ${\mathbf r} = (r_{AC}, r_{AG}, r_{AT}, r_{CG}, r_{CT}, r_{GT})$ \\
Nucleotide frequencies     & $\mbox{\boldmath$\pi$\unboldmath} = (\pi_A, \pi_C, \pi_G, \pi_T)$ \\
Gamma-shape parameter      & $\alpha$
\end{tabular}
\end{center}

We estimate parameters of the phylogenetic model in a Bayesian framework, in which inference is focused on the joint posterior probability distribution of the parameters.
The posterior probability is calculated using Bayes's theorem as
\begin{align}
f(\tau, {\mathbf p}, T, {\mathbf r}, \mbox{\boldmath$\pi$\unboldmath}, \alpha \given {\mathbf X}) = {
f({\mathbf X} \given \tau, {\mathbf p}, T, {\mathbf r}, \mbox{\boldmath$\pi$\unboldmath}, \alpha)
f(\tau, {\mathbf p}, T, {\mathbf r}, \mbox{\boldmath$\pi$\unboldmath}, \alpha)
\over f({\mathbf X})}
\end{align}
where ${\mathbf X}$ represents the alignment(s) of DNA sequences (see below).
Bayes's theorem states that the posterior probability distribution of the parameters [$f(\cdot \given {\mathbf X})$] is equal to the likelihood [$f({\mathbf X} \given \cdot)$]
times the prior probability distribution of the parameters [$f(\cdot)$] divided by the marginal likelihood [$f({\mathbf X})$].
The likelihood is calculated under the phylogenetic model using numerical methods first described by \citet{felsenstein81}.

Bayesian inference treats parameters as random variables; this requires that we specify a prior probability distribution for each model parameter that describes the nature of its random variation.
The prior reflects the biologist's beliefs about the parameters before evaluating the data at hand.
In this study, we assume the following prior probability distributions for the parameters of the phylogenetic model:
\begin{eqnarray*}
\tau & \sim & \mbox{Discrete Uniform}(1, \ldots, B(S)) \\
{\mathbf p} & \sim & \mbox{Dirichlet}(1, 1, \ldots, 1) \\
T & \sim & \mbox{Gamma}(2S-3, \lambda_1) \\
{\mathbf r} & \sim & \mbox{Dirichlet}(1,1,1,1,1,1) \\
\mbox{\boldmath$\pi$\unboldmath} & \sim & \mbox{Dirichlet}(1,1,1,1) \\
\alpha & \sim & \mbox{Exponential}(\lambda_2)
\end{eqnarray*}
where $B(S)=(2S-5)!!$ is the number of possible binary, unrooted trees for $S$ species.
Following \citet[][see also Huelsenbeck and Suchard, 2007]{zhang12}, we specify a compound Dirichlet prior on branch-length proportions, which avoids the so-called `long-tree' problem \citep[{\it sensu} ][]{brown09,rannala12}.

Inference of phylogeny is increasingly based on datasets comprising gene regions sampled from multiple loci and/or different genomes that may have evolved under qualitatively different evolutionary processes. 
How should we accommodate potential heterogeneity in the parameter values within theses datasets?
The program \verb!MrBayes! \citep{ronquist12} addresses this problem by allowing the user to designate subsets of sites within the sequence alignment ({\it i.e.}, `data partitions').
Parameters of the phylogenetic model can then be independently estimated ({\it i.e.}, `unlinked') for individual data partitions, or arbitrarily shared ({\it i.e.}, `linked') across two or more data partitions.
The specific configuration of (un)linking parameters across the specified data partitions corresponds to a specific `partition scheme' ({\it i.e.}, a `mixed model' or `process partition').

Imagine, for example, that our alignment includes a single protein-coding gene.
We define three data partitions for this alignment corresponding to subsets of sites for each of the codon positions.
We now consider how to model the overall substitution rate among the three data partitions.
We might specify a common (completely linked) tree-length parameter for all three data partitions, which reflects the biological assumption that the overall substitution rate does not vary among the three codon positions.
Alternatively, we could specify an independent (completely unlinked) tree-length parameter for each data partition, which assumes that the substitution rate differs among the three codon positions.
Or we could partially link tree-length parameters among data partitions: {\it e.g.}, we might specify a shared tree-length parameter for first- and second-position sites, and an independent tree-length parameter parameter for third-position sites.
The same considerations apply to other parameters in the model (the stationary frequencies, exchangeability rates, etc.).
The difficulty with this approach, however, is to find the best scheme among a bewildering number of alternative ways of linking and unlinking parameters across partitions.

In this study, we use a DPP model to treat the partition scheme as random variable.
This approach obviates the need to condition phylogenetic inference on any individual partition scheme.
Instead, parameters of the phylogenetic model are integrated over all possible partition schemes, visiting each according to its ability to accommodate the underlying process heterogeneity in the data.
Specifically, we invoke an independent DPP model for each of the phylogenetic model parameters---the tree length, exchangeability parameters, base frequencies, and gamma-shape parameter---to describe the process heterogeneity among the pre-specified data partitions. 
The remaining phylogenetic model parameters---the tree topology and branch-length proportions---are assumed to be shared across all of the data partitions.
In the following, we describe the details of this model and the analyses performed in this study.

\begin{center}
{\it Data Partitions}
\end{center}

We assume that the biologist has correctly aligned the nucleotide sequences sampled from $S$ species, and then identified subdivisions of this alignment and partitioned these data accordingly. 
The data partitions might reflect knowledge of the structure of the gene, dividing the data, for example, according to coding versus non-coding DNA, stem versus loop regions of ribosomal genes, or by codon position for protein-coding genes.
For genomic data, the partitions might correspond to individual genes. 
In the most extreme case, the biologist might assign every site to its own subset.
Specification of data partitions is up to the biologist, and becomes a fixed assumption of the inference.

Consider, for example, the following alignment of $S=5$ protein-coding sequences:
\begin{center}
\begin{tabular}{ll}
Human      & {\tt CTGACTCCTGAGGAGAAGTCTGCCGTTACT...} \\
Cow        & {\tt CTGACTGCTGAGGAGAAGGCTGCCGTCACC...} \\
Mouse      & {\tt CTGACTGATGCTGAGAAGTCTGCTGTCTCT...} \\
Marsupial  & {\tt TTGACTTCTGAGGAGAAGAACTGCATCACT...} \\
Chicken    & {\tt TGGACTGCTGAGGAGAAGCAGCTCATCACC...}
\end{tabular}
\end{center}
which are the first $30$ sites of five of the sequences of a partial alignment of $\beta$-globin DNA sequences \citep{yang00}.
This alignment can be partitioned by codon-position, which would result in $K=3$ data subsets, which
we label ${\mathbf X}_1$, ${\mathbf X}_2$, and ${\mathbf X}_3$:
$$
\begin{array}{ccc}
{\mathbf X}_1 = \left( \begin{array}{c}
{\tt CACGGATGGA} \ldots \\
{\tt CAGGGAGGGA} \ldots \\
{\tt CAGGGATGGT} \ldots \\
{\tt TATGGAATAA} \ldots \\
{\tt TAGGGACCAA} \ldots \\
\end{array} \right) 
&
{\mathbf X}_2 = \left( \begin{array}{c}
{\tt TCCAAACCTC} \ldots \\
{\tt TCCAAACCTC} \ldots \\
{\tt TCACAACCTC} \ldots \\
{\tt TCCAAAAGTC} \ldots \\
{\tt GCCAAAATTC} \ldots \\
\end{array} \right) 
&
{\mathbf X}_3 = \left( \begin{array}{c}
{\tt GTTGGGTCTT} \ldots \\
{\tt GTTGGGTCCC} \ldots \\
{\tt GTTTGGTTCT} \ldots \\
{\tt GTTGGGCCCT} \ldots \\
{\tt GTTGGGGCCC} \ldots \\
\end{array} \right) 
\end{array}
$$
for the first, second, and third codon partitions, respectively.
The complete alignment is denoted ${\mathbf X} = ({\mathbf X}_1,{\mathbf X}_2,{\mathbf X}_3)$.

\bigskip

\begin{center}
{\it Combinatorics of Partitions}
\end{center}

The parameters of the phylogenetic model can be assigned to data subsets in a large number of possible ways.
Consider, as an example, a simple case in which the data have been divided into three subsets and a single parameter, $\theta$, is to be assigned to the various subsets.
The parameter can be constrained to be the same in all three subsets (which is equivalent to subdividing the data, and then ignoring the subsets when estimating parameters), allowed to be potentially different in each subset, or constrained to be equal among some subsets but different in others.
The table, below, enumerates all of the possible process partitions for the three data subsets:
\begin{center}
\begin{tabular}{ccccc}
${\mathbf X}_1$ & ${\mathbf X}_2$ & ${\mathbf X}_3$ & & RGF \\ \hline
$\theta_1$ & $\theta_1$ & $\theta_1$ & $\rightarrow$ & (1,1,1) \\
$\theta_1$ & $\theta_1$ & $\theta_2$ & $\rightarrow$ & (1,1,2) \\
$\theta_1$ & $\theta_2$ & $\theta_1$ & $\rightarrow$ & (1,2,1) \\
$\theta_1$ & $\theta_2$ & $\theta_2$ & $\rightarrow$ & (1,2,2) \\
$\theta_1$ & $\theta_2$ & $\theta_3$ & $\rightarrow$ & (1,2,3) \\
\end{tabular}
\end{center}
where the parameter subscript indicates the differentially estimated parameters and the partition is also described according to the restricted growth function (RGF) notation \citep{stanton86}. 
The total number of partitions for $K$ elements is described by the Bell numbers \citep{bell34}.
The Bell number for $K$ elements is the sum of the Stirling numbers of the second kind:
\begin{align} \label{eq:bell_num}
{\cal B}(K) = \sum_{i=1}^K {\cal S}(K,i),
\end{align}
where the Stirling numbers of the second kind are given by the following equation:
\begin{align}
{\cal S}(n, k) = {1 \over k!} \sum_{i=0}^{k-1} (-1)^i {k \choose i} (k-i)^{n}.
\end{align}

\bigskip

\begin{center}
{\it Dirichlet Process Prior Model}
\end{center}

The Dirichlet Process Prior model is a probability model on partitions \citep{ferguson73,antoniak74}.
This model contains a singe parameter, called the concentration parameter, $\chi$, that controls how probability mass is assigned to different partitions.
For the simple example in which there are $K=3$ data subsets, the DPP model distributes probability to each process partition:
\begin{center}
\begin{tabular}{cc}
RGF     & Prob. \\ \hline
(1,1,1) & $1/3$ \\
(1,1,2) & $1/6$ \\
(1,2,1) & $1/6$ \\
(1,2,2) & $1/6$ \\
(1,2,3) & $1/6$ \\
\end{tabular}
\end{center}
(For this example, we assume $\chi = 1$.)
More generally, the probability of a particular partition of $K$ elements containing $k$ subsets is
\begin{align}
p({\mathbf z}, k \given \chi, K) = \chi^k { \prod_{i=1}^k(\eta_i - 1)! \over \prod_{i=1}^K (\chi + i -1) },
\end{align}
where ${\mathbf z}$ is the allocation vector specifying how elements are assigned to subsets of the partition, $k$ is the number of subsets in the partition (or the `degree' of the partition), and $\eta_i$ is the number of elements assigned to subset $i$. For the case in which $K=3$ the values for these parameters are
\begin{center}
\begin{tabular}{ccccc}
RGF     & ${\mathbf z}$ & $K$ & $k$ & $\eta_{\cdot}$ \\ \hline
(1,1,1) & (1,1,1) & 3 & 1 & $\eta_1=3$ \\
(1,1,2) & (1,1,2) & 3 & 2 & $\eta_1=2$, $\eta_2=1$ \\
(1,2,1) & (1,2,1) & 3 & 2 & $\eta_1=2$, $\eta_2=1$ \\
(1,2,2) & (1,2,2) & 3 & 2 & $\eta_1=1$, $\eta_2=2$ \\
(1,2,3) & (1,2,3) & 3 & 3 & $\eta_1=1$, $\eta_2=1$, $\eta_3=1$ \\
\end{tabular}
\end{center}

The DPP model contains one additional component: a probability distribution (sometimes referred to as the `base distribution', denoted $G_0$) for the parameter assigned to each partition.
A complete description of the process, then, specifies the concentration parameter and the base distribution for the parameter.
The DPP model is often used in Bayesian analysis for clustering.
The elements of the partition are the objects to be clustered.
Typically, the objects to be clustered correspond to data elements.
The subsets of the partition denote the clustering; the subsets specify which elements are grouped together in the same cluster.
Importantly, under the DPP model the number of clusters is a random variable.
The DPP model does not force the biologist to pre-specify the number of clusters {\it a priori}.

The Chinese Restaurant Process is a useful metaphor that has been used to describe the DPP model \citep{aldous85}.
The metaphor works as follows: Consider a Chinese restaurant with a countably infinite number of tables.
Patrons enter the restaurant one at a time and randomly choose a table at which to sit.
The $i$th patron will choose to sit at table $m$ at which $\eta_m$ patrons are already seated with probability ${\eta_m \over i-1+ \chi}$; alternatively, this patron will choose some unoccupied table with probability ${\chi \over i-1+ \chi}$.
The `restaurant patrons' in the metaphor are replaced by data elements.
Data elements that are `seated at the same table' share a common parameter, $\theta$, which is drawn from the probability distribution $G_0$ ({\it i.e.}, $\theta \sim G_0$).
Note that tables with large numbers of patrons will tend to attract new patrons.

We apply a separate DPP model to each of the phylogenetic model parameters, except for the topology and branch lengths of the tree \citep[however, see][]{ane07}.
Specifically, we first identify process partitions in the data and then allow the specific phylogenetic model parameters---the nucleotide frequencies, substitution rates, tree length, and gamma shape parameter---to cluster across partitions according to a DPP model.

\bigskip

\begin{center}
{\it Markov Chain Monte Carlo}
\end{center}

We use Markov chain Monte Carlo \citep[MCMC;][]{metropolis53,hastings70} to approximate the posterior probability distribution of the phylogenetic parameters.
The general idea is to construct a Markov chain that has as its state space the possible values for the model parameters and a stationary probability distribution that is the target distribution of interest ({\it i.e.}, the posterior probability distribution of the model parameters).
Draws from the Markov chain at stationarity are valid, albeit dependent, samples from the posterior probability distribution \citep{tierney94}.
Accordingly, parameter estimates are based on the frequency of samples drawn from the stationary Markov chain.

MCMC works by iterating the following steps many thousands or millions of times: 
(1) Call the current state of the Markov chain $\theta$.
If this is the first cycle of the Markov chain, then initialize $\theta$ to some value (perhaps by drawing values from the corresponding prior probability distribution).
(2) Propose a new value for the model parameter $\theta$, called $\theta'$.
The proposal mechanism must be stochastic, with a probability of $f(\theta \rightarrow \theta')$ of choosing the proposed state $\theta'$ from the current state $\theta$.
The probability of the reverse proposal, from $\theta'$ to $\theta$, which is not actually made in computer memory, is $f(\theta' \rightarrow \theta)$. 
(3) Calculate the probability of accepting $\theta'$ as the next state of the Markov chain:
\begin{align} \label{eq:acceptance_prob}
R = \min \left(1, {\Pr(\mathbf{X} \given \theta') \over \Pr(\mathbf{X} \given \theta)} \times {\Pr(\theta') \over \Pr(\theta)} 
\times {f(\theta' \rightarrow \theta) \over f(\theta \rightarrow \theta')} \right).
\end{align}
(4) Generate a uniform(0,1) random variable called $u$. If $u < R$, then accept the proposed state and set $\theta = \theta'$. 
(5) Go to step 1.

We update the parameters of the phylogenetic model one-at-a-time.
The proposal mechanisms for most of the phylogenetic model parameters, such as the tree topology, nucleotide frequencies, exchangeability parameters, and gamma-shape parameter, are now rather well established \citep[{\it e.g.},][]{larget99,holder05,lakner08}.
However, the proposal mechanism for the partitions with a Dirichlet process prior model merit further discussion.
\citet{neal00} discussed several possible MCMC updates for DPP models.
We used `Algorithm 8' from that paper to update the partition schemes. 
To return to the Chinese Restaurant metaphor, we imagine that we have instantiated a restaurant in computer memory, with tables
representing subsets of the partition scheme and patrons who are seated at tables representing the sites (or, as currently implemented, sets of sites that
are determined by the biologist's initial division of the data into process partitions).

The MCMC procedure works as follows. 
First, pick a patron and remove it from the table. If the patron was the only individual at that table, then remove the table from computer memory.
Otherwise, decrease the count of the number of patrons at the table by one ({\it e.g.}, decrease $\eta_i$).
Calculate the likelihood when the patron is seated at 
the $m$th remaining table in computer memory ($L_m$).
Also, calculate the likelihood when the patron is seated at each of the $\kappa$ `auxiliary' tables.
These auxiliary tables are instantiated for the purpose of allowing the patron to be seated at a new (unoccupied) table.
The parameter values for the auxiliary tables are drawn from the prior probability distribution for the parameter.
The probability of seating the patron at the $m$th table in computer memory at which $\eta_m$ patrons are already seated is
$C \times \eta_m \times L_m$, and the probability that the patron is seated alone at the $k$th auxiliary table is $C \times (\chi/ \kappa) \times L_k$, where
$C$ is a normalizing constant.
After the patron is reseated at a previously existing or a new (auxiliary) table, the unoccupied auxiliary tables are deleted from computer memory.
One MCMC cycle involves a scan of all the patrons, with the above update mechanism applied to each.

The methods described in this paper---which are intended to automate the phylogenetic analysis of partitioned datasets---have been implemented in the freely available
application, \verb!AutoParts!.
This command-line program enables a fully hierarchical Bayesian phylogenetic analysis of partitioned data, where the input is an alignment of nucleotide sequences and two or more pre-specified data subsets.
The program uses MCMC to approximate the marginal posterior probability distribution of all parameters (tree topology, branch lengths, and substitution model parameters) while simultaneously integrating over all possible partition schemes under the DPP model.

\newpage

\begin{center}
{\it Summarizing Partitions}
\end{center}

In a Bayesian MCMC analysis, posterior probabilities of model parameters are based upon the MCMC samples. 
For example, the posterior probability of a particular phylogenetic tree is approximated by the fraction of the time that the Markov chain visited that particular tree topology. 
One of the challenges of MCMC-based phylogeny estimation is how to most appropriately summarize the information within the MCMC samples.
For some simple model parameters, such as  $\alpha$---the parameter specifying the shape of the gamma distribution used to describe among-site rate variation---we can simply summarize the corresponding estimate by constructing a frequency histogram of the sampled $\alpha$ values; this histogram is an approximation of the marginal posterior probability density.
The marginal posterior probability density can be further summarized in various ways, {\it e.g.}, as the mean of the posterior density, or an interval containing $95\%$ of the posterior probability ({\it i.e.},  the $95\%$ credible interval).
By contrast, other parameters---such as the tree topology, $\tau$---presents some difficulties in a MCMC analysis, as it is an unusual model parameter \citep[][]{yang95b}.
Consequently, there is no natural way to summarize an MCMC sample of tree topologies, except perhaps to simply report the posterior probabilities of individual trees.
Nevertheless, several useful methods have emerged as conventions for summarizing a posterior sample of trees.
For example, the majority-rule consensus tree is often used to summarize the results of an MCMC analysis \citep[in the same way that bootstrap samples are summarized;][]{felsenstein85a}.

Our MCMC procedure also samples the process partitions of the model parameters.
Partitions, like trees, are not standard model parameters and it is not obvious how the information contained in an MCMC sample of partitions should be summarized.
One possibility is to simply report the posterior probabilities of individual partitions.
This method, however, is impractical when the posterior probability is spread across many partitions and the Markov chain may then explore many thousands or millions of partitions.
In this study, we summarize the MCMC samples of partitions using the concept of a `mean partition'. 
The mean partition, denoted $\bar{\sigma}$, is the partition that minimizes the sum of the squared distances to all of the sampled partitions.
We use a distance on partitions first described by \citet{gusfield02}.
The Gusfield distance between two partitions can be interpreted as the minimum number of elements that must be removed from both partitions to make the induced partitions identical. 
For example, the partitions $\sigma_1 = (1,1,1,1,2,2)$ and $\sigma_2 = (1,1,1,2,2,2)$ can be rendered identical by removing the fourth element resulting in the induced partition $(1,1,1,2,2)$; in this example, the partition distance is $d_p = 1$.

\bigskip

\newpage
\begin{center}
{\it Analysis of Empirical Data}
\end{center}
{\it Sequence alignments and data partitions.}---We demonstrate the DPP approach by applying it to three published datasets.
Our choice of datasets was motivated by two considerations.
First, these datasets have previously been the focus of careful analysis using the conventional mixed-model selection approach (where Bayes factors are used to choose among candidate partition schemes), and so they provide a nice point of comparison to the DPP approach described here.
Second, these datasets span a range of size (in terms of the number of species, $S$, and sequence length, $L$) and number of data subsets, $K$, which provides an opportunity to assess the ability of the new method in a variety of empirical settings.
For each dataset, we assume an identical sequence alignment and adopt the same data subsets as those used in the original studies.

First, we analyzed the alignment of $S = 32$ gall wasp sequences from \citet{nylander04}.  
The gall wasp alignment includes $L = 3080$ sites, which we partitioned into $K = 11$ data subsets.
Specifically, we partitioned each of the three protein-coding genes---including the two nuclear genes (elongation factor 1$\alpha$ [EF1$\alpha$] and long-wavelength opsin [LWRh]) and the mitochondrial gene region (cytochrome oxidase {\it c } subunit I [COI])---by codon position. 
We also partitioned the 28S ribosomal sequence by stem and loop regions.  

Second, we analyzed an alignment of $S = 60$ skink sequences from \citet{brandley05}.  
The skink alignment includes $L = 2678$ sites, which we partitioned into $K = 11$ data subsets.
Specifically, we partitioned the mitochondrial gene region (NADH dehydrogenase subunit 1; ND1) by codon position.
We also specified individual data partitions for the three tRNA sequences: tRNA\textsuperscript{GLU}, tRNA\textsuperscript{ILE}, and tRNA\textsuperscript{GLN}.
Finally, we specified individual data partitions for the three ribosomal sequences: 12S rRNA, 16S rRNA, and tRNA\textsuperscript{MET}.  

Third, we analyzed an alignment of $S = 164$ hummingbird and relevant outgroup sequences from \citet{mcguire07}.  
The hummingbird alignment includes $L = 4122$ sites, which we partitioned into $K = 9$ data subsets.
Specifically, we partitioned each of the three protein-coding genes---including the two mitochondrial gene regions (NADH dehydrogenase subunit 2 [ND2] and partial NADH dehydrogenase subunit 4 [ND4])---by codon position.
We also specified a separate partition for the tRNA flanking regions.
Finally we specified two individual data partitions for each of the nuclear sequences---including intron $5$ of the adenylate kinase gene (AK1) and intron $7$ of the beta fibrinogen (BFib) gene regions.  

\bigskip
\noindent
{\it Exploring the impact of the concentration parameter on phylogenetic inferences.}---The DPP approach for accommodating process heterogeneity introduces a single novel parameter, the concentration parameter, $\chi$, that controls the expected number of process partitions, $E(k)$.
In order to assess the sensitivity of posterior estimates to the specified prior on $\chi$, we performed two series of analyses.
Our first series of analyses fixed the concentration parameter, $\chi$, to a set of values.
We chose values for $\chi$ that centered prior mean on the number of process partitions, $E(k)$, to a corresponding set of values that spanned the possible range for each dataset.  
For example, in our analyses of the gall wasp and skink alignments, which both have $K = 11$ data subsets, we fixed the concentration parameter, $\chi$, to a set of values such that the prior mean for the number of process partitions spanned the entire range of possible values---specifically such that $E(k) = \{\approx 1.0, 2.0, 4.0, 6.0, 8.0, 10.0, \approx 11.0\}$.
Similarly, our analyses of the hummingbird alignment, with $K = 9$ data subsets, specified a set of values for $\chi$ such that $E(k) =\{\approx 1.0, 2.0, 4.0, 6.0, 8.0, \approx 9.0\}$.
We performed independent analyses under each of the corresponding values of $E(k)$ to assess the sensitivity of posterior estimates to the prior on the concentration parameter, $\chi$.

Our second series of analyses explored an identical range of values for $E(k)$, but in these analyses we treated the concentration as a random variable.
Our implementation of the DPP model allows the concentration parameter, $\chi$, to be fixed to a specific value [which will center $E(k)$ to a corresponding value], or to use a hierarchical Bayesian modeling approach that treats the concentration parameter as a random variable.
Specifically, we allow the concentration parameter, $\chi$ to be modeled as a gamma-distributed random variable: under this option, it is necessary only to specify the value of the parameter controlling the shape of the gamma hyperprior, such that $\chi$ is estimated from the data.
As in the fixed-$\chi$ specification of the DPP model, the mean of the gamma hyperprior on $\chi$ can be specified to center the prior mean number of process partitions, $E(k)$, to a desired value.
In this second series of analyses, we specified the gamma hyperprior on $\chi$ to a set of values that centered the prior mean on the number of partitions, $E(k)$, to the same range of values as those we explored in the first series of analyses ({\it i.e.}, where the concentration parameter was treated as fixed).

\bigskip
\noindent
{\it MCMC analyses.}---We estimated the joint posterior probability distribution for each of the three study groups under the Dirichlet process prior model using the Markov chain Monte Carlo (MCMC) algorithms implemented in \verb!AutoParts!.
Each MCMC simulation was initiated from random parameter values (drawn from the corresponding priors), run for 2x10$^7$ cycles, and thinned by sampling every $1000$\textsuperscript{th} state of the chain.
In total, we performed $480$ separate MCMC analyses on the empirical datasets: we analyzed each of the three alignments under the appropriate range of values for the expected number of process partitions, $E(k)$, and for each value of $E(k)$ we performed replicate analyses where the concentration parameter was treated first as fixed and then as a random variable (as described above), and performed four replicate analyses under each uniquely specified analysis (to help assess MCMC performance).  

\bigskip
\noindent
{\it Assessing MCMC performance.}---Convergence of each chain to the stationary distribution was assessed by plotting time series of sampled values for all continuous parameters (exchangeability rates, stationary frequencies, tree length, alpha-shape parameter) using \verb!Tracer! v.1.5 \citep{rambaut07}.
We also assessed convergence by comparing independent estimates of the marginal posterior probability density for each continuous parameter, ensuring that parameter estimates from the four independent MCMC simulations were effectively identical and SAE compliant \citep[{\it c.f.},][]{Brooks_1997}.
Similarly, we assessed convergence by comparing independent estimates of the discrete model parameter---the tree topology---using the `{\it comparetree}' function implemented in \verb!MrBayes! v.3.1.2 \citep{ronquist03} to assess the correspondence of clade posterior probabilities estimated by pairs of independent chains. 
We assessed the efficiency (`mixing') of the MCMC simulations both by monitoring the acceptance rates of proposals for all parameters, and by computing the estimated sample size (ESS) diagnostic implemented in \verb!Tracer! v.1.5 \citep{rambaut07}.

\bigskip
\noindent
{\it Validating the MCMC algorithms.}---We performed two series of experiments to validate the MCMC algorithms implemented in \verb!AutoParts!.
First, we assessed our ability to recover the prior.
If an MCMC simulation is run without data, the likelihood for all parameter values will be identical.
Therefore, the ratio of likelihoods for proposed and current states of the chain will equal one.
Accordingly, acceptance probabilities of sequential states of the chain will be determined entirely by the prior ratio of proposed and current states (\emph{c.f.}, eqn. \ref{eq:acceptance_prob}).
Consequently, a valid MCMC algorithm run without data will target the joint \emph{prior} probability density rather than the joint \emph{posterior} probability density.
The \emph{estimated} marginal prior probability density for each parameter can then be compared to its \emph{known} (specified) counterpart.
We used this approach on simulated and empirical datasets, exploring the ability of the MCMC implementation in \verb!AutoParts! to correctly recover the prior over a range of values for the concentration parameter.  

We also validated the MCMC algorithms---and provided a benchmark for inference under the DPP model---by estimating the posterior probability distributions for each of the three study groups under a uniform GTR+$\Gamma$ substitution model using \verb!AutoParts! and the proven Metropolis-coupled MCMC (or MC$^3$) algorithms implemented in \verb!MrBayes! v.3.1.2 \citep{ronquist03}. 
For the unpartitioned analyses using \verb!MrBayes!, we enforced a uniform GTR+$\Gamma$ substitution model, where each parameter was estimated from all of the sites comprising a given sequence alignment.
For the unpartitioned analyses using \verb!AutoParts!, we enforced a uniform GTR+$\Gamma$ substitution model by setting the proposal probability of new partitions to zero, which prevented the MCMC from considering partition schemes of ${\it k} >1$.  

To assess the ability of \verb!AutoParts! to recover the prior, we repeated the $480$ MCMC analyses of the empirical data described above, in this case targeting the joint prior (rather than posterior) distributions.
We performed an additional $24$ MCMC simulations for the benchmark analyses.
Specifically, we ran four independent MCMC simulations for each of the three alignments, and for each alignment we performed analyses under a uniform GTR+$\Gamma$ substitution model using both \verb!AutoParts! and \verb!MrBayes!.
We performed each of the $504$ MCMC simulations as described above.
However, our analyses using the MC\textsuperscript{3} algorithm in \verb!MrBayes! were performed using four incrementally heated chains (where the parameter governing the `temperature' of the heated chains ranged from ${0.2 - 0.1}$).
We assessed performance of the MCMC simulations as described previously; however, we also assessed convergence of the \verb!MrBayes! analyses by running `paired'-MC\textsuperscript{3} simulations ({\it i.e.}, two independent, parallel simulations using four incrementally heated chains) and monitoring the average standard deviation of split frequencies sampled by the paired cold chains.

\begin{center}
{\it Analysis of Simulated Data}
\end{center}

\noindent
{\it Simulation study design.}---To complement the empirical demonstration of the DPP approach, we also assessed the ability of this method to successfully recover parameter values under simulation.
Simulation is a critical tool for validating an inference method, as it can uniquely assess the ability of an approach to recover the true/known parameter values that were used to generate the data.
The success of a model on simulated data, however, may not provide a reliable guide to its empirical behavior: the design of the simulation study therefore warrants careful consideration.
A simulation study will provide meaningful insights into the statistical behavior of a given method only if it adequately explores the relevant parameter space.
Conceptually, this entails defining an $N$-dimensional hypervolume (for methods impacted by $N$ variables, such as tree size, number and size of data partitions, number of process partitions, degree of process heterogeneity, etc.) that encompasses conditions that are likely to be encountered when the method is applied to real data.
This space is first discretized into a number of cells (each corresponding to a unique parameter combination), and each cell is then explored with an adequate intensity (by simulating a sufficient number of realizations under each parameter combination).
Each of the simulated datasets is then analyzed using the inference to assess how its statistical behavior varies over parameter space. 
The dimensionality of this parameter space and the associated computational burden can quickly become immense, particularly for complex models.
	
\begin{figure}[h] 
\centering 
\includegraphics[width=90mm]{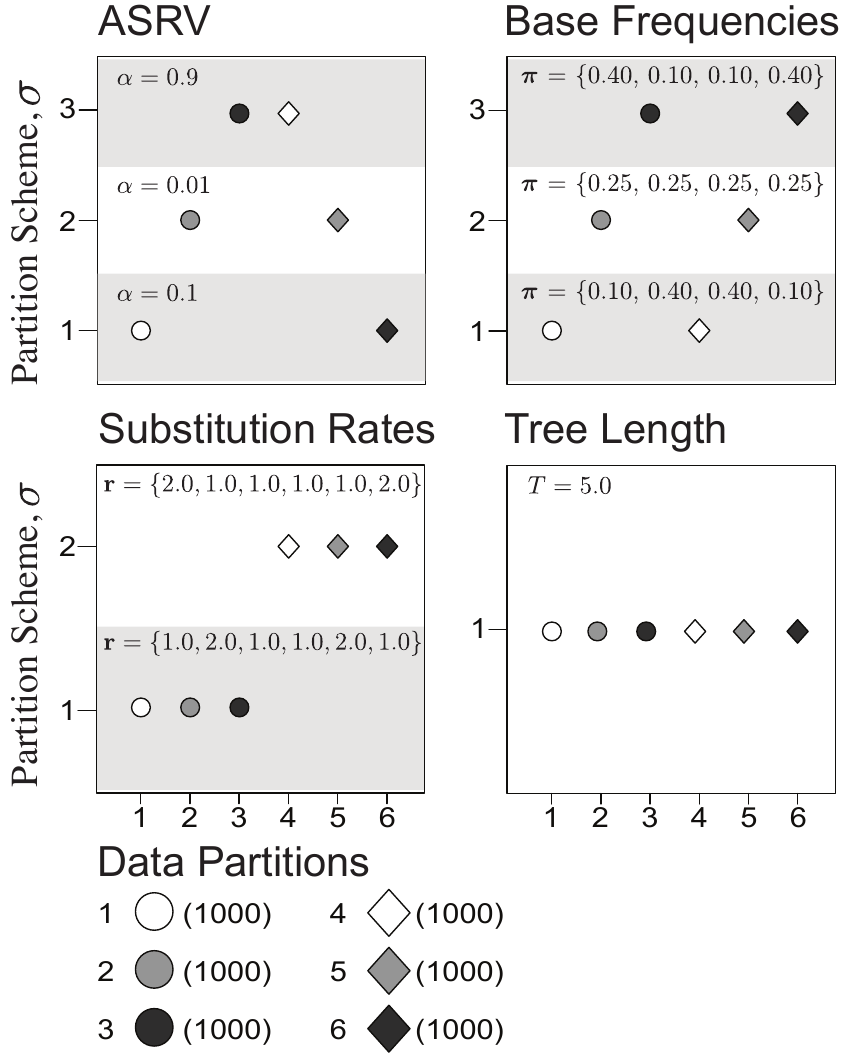} 
\caption{A typical simulation experiment.  This simulation involves $K = 6$ data partitions, each with $L = 1000$ sites. All data partitions are assumed to share a common tree topology, $\tau$, and branch-length proportions, ${\mathbf p}$, but other aspects of the substitution process may differ between data partitions, including: the degree of among-site rate variation (ASRV, governed by the $\alpha$-shape parameter); the vector of nucleotide frequencies, \mbox{\boldmath$\pi$\unboldmath}; the vector of exchangeability rates, ${\mathbf r}$, and the overall rate of substitution/tree length, $T$.  The partition scheme, $\sigma$, for each of the four parameters specifies both the number of process partitions, $k$, and the allocation vector, ${\mathbf z}$, of the $K = 6$ data partitions to those $k$ process partitions. For example, the 6 data partitions are allocated to $k = 3$ distinct process partitions for the $\alpha$-shape parameter (the shaded rows of the panel), where the allocation vector is ${\mathbf z}  = (1,2,3,3,2,1)$ (depicted by the assignment of the data-partition symbols to their respective rows), and the specific parameter values for each process partition are indicated in the corresponding rows of the panel.}
\label{sim_design}
\end{figure}

Accordingly, a comprehensive simulation study exploring the statistical behavior of the DPP approach is beyond the scope of the current study, and will be published separately (Moore and Huelsenbeck, in prep.).  
Herein, we present a simulation experiment that is typical both of the design of the more general simulation study and also of the statistical behavior of the DPP approach. 
Specifically, we simulated $100$ datasets with $K=6$ data partitions (each with $L = 1000$ sites) for a tree with $S=50$ species.
We assumed that the six data partitions share a common tree topology, $\tau$, and branch-length proportions, ${\mathbf p}$, but allowed other aspects of the substitution process to differ among data partitions.
For each simulation, the tree topology, $\tau$, and vector of branch-length proportions, ${\mathbf p}$, were drawn from their corresponding prior probability distributions.
Values for the other substitution model parameters were specified as described in Figure \ref{sim_design}.  
We specified values for the concentration parameter, $\chi$, following the protocol used in the empirical analyses.
Specifically, we assessed the sensitivity of estimates to the prior on $\chi$ (and the treatment of $\chi$ as a fixed or random variable)
by analyzing each simulated dataset under the possible range of values for the expected number of process partitions, $E(k)=\{\approx 1.0, 3.0, \approx 6.0\}$, which we specified using both fixed values of $\chi$ and also using a gamma hyperprior on $\chi$.
Careful simulation represents a non-trivial enterprise: the simulation experiment presented here consumed approximately $403,200$ hours ($\approx 46$ years) of compute time. 

\bigskip
\noindent
{\it MCMC analyses and performance assessment.}---We estimated the posterior probability distribution of parameters for each simulated dataset under the Dirichlet process prior model using \verb!AutoParts!.
Each MCMC simulation was initiated from random parameter values (drawn from the corresponding priors), run for 2x10$^6$ cycles, and thinned by sampling every $1000$\textsuperscript{th} state of the chain.
In total, we performed $1800$ separate MCMC analyses on the simulated datasets: we analyzed each of the $100$ replicate datasets under the possible range of values for the expected number of process partitions, $E(k)=\{\approx 1.0, 3.0, \approx 6.0\}$, and for each value of $E(k)$ we performed replicate analyses where the concentration parameter was treated first as fixed and then as a random variable, and performed three replicate MCMC simulations under each uniquely specified analysis (to help assess MCMC performance).  
We assessed convergence and mixing of each MCMC analysis of the simulated datasets following the protocol described previously for the empirical analyses. 

We assessed the statistical performance of the DPP approach based on properties of the $95\%$ credible intervals (CI) of process partitions estimated for the simulated datasets.  
Specifically, we computed the $95\%$ CI of process partitions for each MCMC sample by first ordering the sampled process partitions by their posterior probability.
Next, we added the probability for the first process partition to that for the second process partition, and then added this sum to the probability for the third process partition, and so on, until the \emph{cumulative} probability was equal to or greater than $0.95$.
The resulting set of (one or more) process partitions corresponds to the $95\%$ CI. 
We then used the $95\%$ CIs on process partitions to assess the \emph{accuracy} and the \emph{precision} of the DPP approach. 
Accuracy is defined by the \emph{coverage probability}, which is simply the proportion of replicates for which the $95\%$ CIs contain the true process partition.
Coverage probabilities $\geq 0.95$ are considered unbiased and robust estimators.
Precision is defined as a function of the size of the $95\%$ CI: for a given coverage probability, estimators with a smaller CI are more precise.
We measured the precision of estimates as a function of the average number of process partitions comprising the $95\%$ CIs, $\overline{\text{CI}}$, relative to the maximum possible CI size, $\text{CI}_{\text{max}}$: 
$$
\text{Precision} = \frac{\text{CI}_{\text{max}}-\overline{\text{CI}}}{{\text{CI}_{\text{max}}}-1} \times 100,
$$
\noindent
where the maximum possible size of the $95\%$ CI is defined as the Bell number for $K$ data partitions, $\text{CI}_{\text{max}} = {\cal B}(K)$ (\emph{c.f.}, eqn. \ref{eq:bell_num}).
The precision therefore ranges from $0-100$, where larger values indicate higher precision.

\bigskip

\begin{center}
{\sc Results}
\end{center}

\noindent
{\it Performance of MCMC algorithms.}---In general, the MCMC algorithms implemented in \verb!AutoParts! to estimate the joint posterior probability density of trees and other parameters under the Dirichlet process prior model appear to have performed well both for the empirical and simulated datasets, whether we enforced uniform substitution models (for the 
purposes of algorithm validation), or when jointly estimating the process partitions with the concentration parameter either fixed to a specified value or governed by a gamma hyperprior.
Validity of the MCMC algorithms is also confirmed by their ability to successfully recover the prior when run without data.

Under baseline inference scenarios---where a uniform GTR+$\Gamma$ substitution model was enforced---independent estimates obtained using the MCMC algorithms implemented in \verb!AutoParts! and the MC$^3$ algorithms implemented in \verb!MrBayes! were virtually identical (see {\it Supplementary Material}).
Congruence of estimated marginal posterior probability densities of all model parameters---including the overall marginal log likelihoods and clade probabilities---suggests that the MCMC algorithms used to implement the Dirichlet process prior model in \verb!AutoParts! performed appropriately under these simplified inference conditions.

Under more complex inference scenarios---where inferences were integrated over all possible partition schemes under the Dirichlet process prior model---the MCMC algorithms again appear to have performed well. 
Time series plots of all model parameters sampled by each of the four independent MCMC analyses under each of the 56 unique concentration-parameter treatments that we explored for the three empirical alignments appeared to stabilize within five million cycles (see {\it Supplementary Material}).
Accordingly, parameter estimates were based on the final 15 million cycles from each of the four independent MCMC simulations.
Parameter estimates for the simulated datasets were conservatively based on the latter half of samples drawn in each MCMC simulation.

Moreover, independent chains inferred virtually identical marginal posterior probability densities for all parameters---including those of the GTR substitution model, the $\alpha$-shape parameter used to model among-site substitution rate variation, as well as the clade posterior probabilities, and the mean process partition scheme for each of the three empirical alignments---providing additional evidence that the MCMC algorithms used to implement the Dirichlet process prior model successfully converged to the stationary distributions for these datasets (see {\it Supplementary Material}). 

The MCMC algorithm appears to have mixed well over the joint posterior densities for the three empirical datasets: acceptance rates of proposal mechanisms for all parameters were within the target window $({20-70\%})$, and the marginal posterior probability densities for all parameters were typically tight and focused.  
Samples drawn during the stationary phase of independent chains were pooled, which provided adequate sampling intensity for all estimated parameters for all three datasets ({\it i.e.}, all ESS values ${>>10^3}$).

\bigskip
\noindent
{\it Performance of the DPP approach under simulation.}---The results of our simulation experiment demonstrate the ability of the method to reliably recover true/known parameter values (Table \ref{tab:simulation}).
In fact, the statistical performance of the DPP approach in this simulation was remarkably high: the $95\%$ credible set of process partitions included the true process partition with an average coverage probability of $1.0$, reflecting essentially perfect accuracy.
Moreover, the average size of the $95\%$ credible set of process partitions was very small (comprising on average $\bar{x} = 2.6$ process partitions), reflecting a very high degree of precision ($\bar{x} = 99.2$).
These results were insensitive both to the value of the concentration parameter ($\chi$, controlling the prior mean on the number of process partitions, $E(k)$), and also to the approach used to specify the  concentration parameter (either as a fixed value or gamma-distributed random variable).
Although these results are encouraging, the statistical behavior of the DPP approach will benefit from a much more comprehensive simulation study (Moore and Huelsenbeck, in prep.).
The remainder of the results focus on empirical applications of the DPP approach.

\bigskip
\noindent
{\it Sensitivity of inferences to the concentration parameter.}---The two primary series of analyses explored alternate approaches for specifying the concentration parameter of the DPP model.  
Whether we fixed the value of $\chi$ or a placed a gamma hyperprior on $\chi$ to center the prior mean on the number of process partitions at a particular value [{\it e.g.}, $E(k)=4$], inferences were virtually indistinguishable both for the simulated datasets (Table \ref{tab:simulation}) and empirical alignments (see {\it Supplementary Material}). 
This result has important practical implications: most empirical applications of the DPP approach will lack a basis for committing to a particular fixed value of $\chi$, which would require several independent analyses that increment $\chi$ over a range of values for $E(k)$.  
However, our findings suggest that a vague hyperprior can be placed on $\chi$ so that it can be reliably estimated from the data (or equivalently, inferences can be integrated over a suitably broad prior distribution for $\chi$) in the course of a single MCMC analysis \citep[{\it c.f.},][]{escobar95, west94, gelfand05, dorazio09}.
Because inferences under the two approaches were virtually identical, the remainder of the discussion will focus exclusively on results obtained under fixed values of $\chi$.

\begin{figure}[h] 
\centering 
\includegraphics[width=120mm]{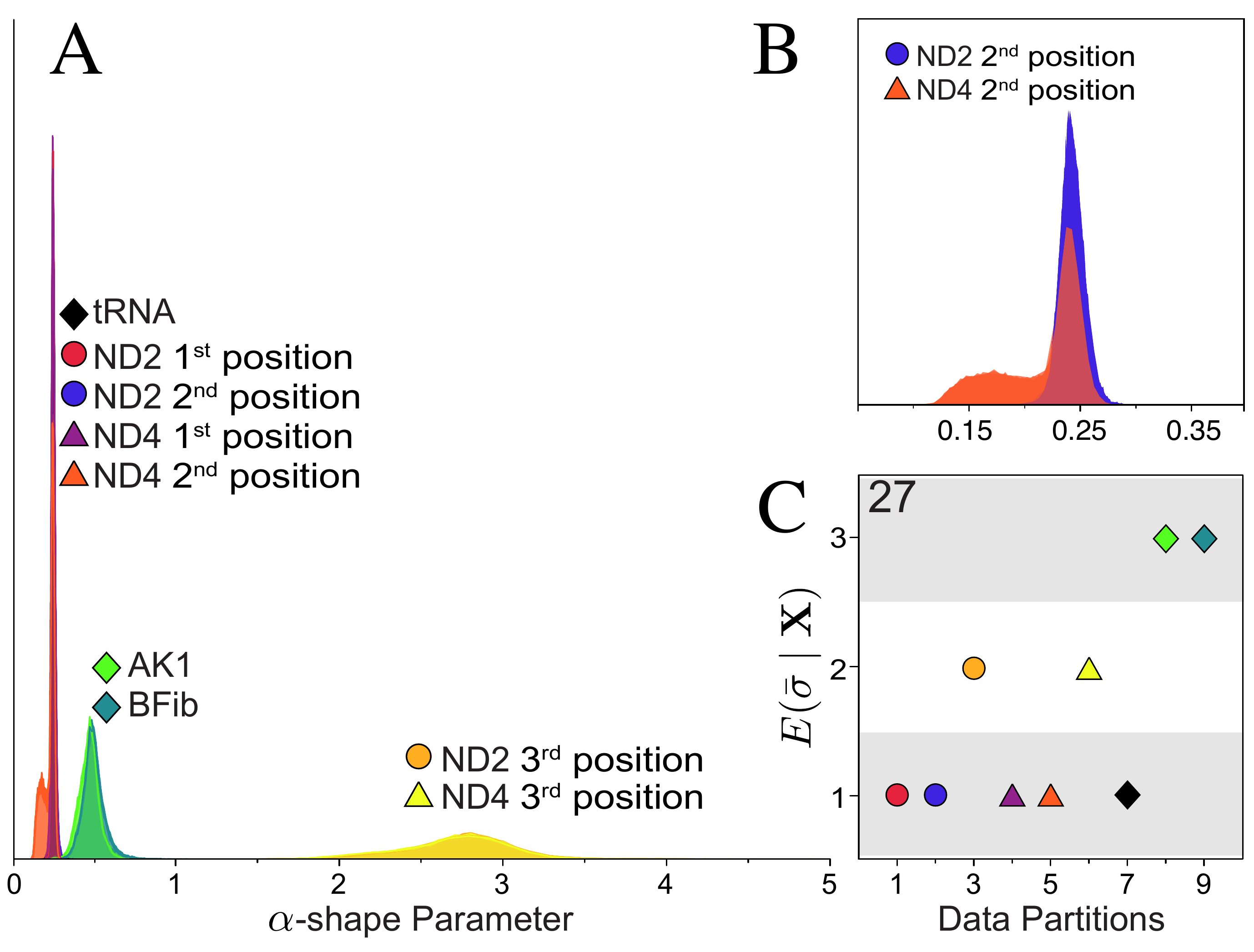} 
\caption{Inferred patterns of among-site substitution rate variation within the hummingbird sequence alignment.  (A) Marginal posterior probability densities for the $\alpha$-shape parameter of the discrete gamma model have been estimated for each of the 9 predefined data partitions.  (B) In contrast to the unimodal marginal densities for other data partitions, the posterior density for the second-position sites of the ND4 gene is distinctly bimodal, suggesting residual process heterogeneity within this data partition.  (C) Analyses under the Dirichlet process prior model provide estimates both for the number of process partitions, $E(k \mid \mathbf{X}) = 3$ (represented as rows in the panel), as well as the mean partition scheme, $E(\bar{\sigma} \mid \mathbf{X})$ (depicted by the assignment of the data-partition symbols to their respective rows).}
\label{results_example}
\end{figure}

\bigskip
\noindent
{\it Some specific results.}---Before describing more general aspects of our findings, we will first focus on results for a particular (albeit fairly typical) case.  Specifically, estimates of the $\alpha$ parameter governing the shape of gamma distribution used to model variation in substitution rates across sites in the 
hummingbird dataset (Figure \ref{results_example}).
The marginal posterior probability density of the $\alpha$-shape parameter is independently estimated for each of the $K = 9$ data partitions (Figure \ref{results_example}A), where the density for each data partition is an amalgam of 24 overlain estimates---that is, the four replicate MCMC analyses performed under each of the six values for the prior mean on the number of process partitions [$E(k) = \{\approx 1.0, 2.0, 4.0, 6.0, 8.0, \approx 9.0\}$] that we explored for the hummingbird dataset.

Four aspects of these results warrant comment.  
First, for each data partition, the marginal posterior probability densities estimated by the four independent MCMC analyses {\it under each particular value} of $E(k)$ are virtually identical, indicating that the independent chains successfully converged to the stationary distributions.

Second, the marginal posterior probability densities of the $\alpha$-shape parameter estimated from each data partition {\it over the entire range of values} for $E(k)$ are equally indistinguishable, suggesting that these parameter estimates are robust to (mis)specification of the concentration parameter of the Dirichlet process prior model.

Third, the marginal posterior probability densities are typically focused and unimodal in form, suggesting both that the MCMC algorithms mixed well over these parameters, and that the pre-specified data partitions successfully captured process heterogeneity in the sequence alignment.  
A notable exception is apparent for the second-position sites of the ND4 gene (Figure \ref{results_example}B): the marginal posterior probability density for this data partition is clearly bimodal, indicating residual heterogeneity in the degree of among-site substitution rate variation within this data partition.  
Because the data partitions are pre-specified by the investigator, the latent structure within the second-position sites of the ND4 gene is not accessible to the Dirichlet process prior model.

Finally, the set of marginal posterior probability densities for the nine data partitions appear to cluster in three distinct groups: the first cluster comprises data partitions for tRNA and the first- and second-position sites of the ND2 and ND4 genes (with relatively high levels of among-site substitution rate variation), the 
second cluster comprises partitions for the AK1 and BFib introns (with intermediate levels of among-site substitution rate variation), and the final cluster includes the two data partitions for the third-position sites of the ND2 and ND4 genes (with relatively low levels of among-site substitution rate variation).  
Indeed, this allocation of data partitions among process partitions is identical to the mean partition scheme identified by the Dirichlet process prior model.  
For convenience, we summarize these results using the graphical convention depicted in Figure \ref{results_example}C.  
This plot reveals the posterior estimate for the number of process partitions, $E(k \mid \mathbf{X}) = 3$, and the inferred mean partition scheme for this parameter, $E(\bar{\sigma} \mid \mathbf{X})$, which summarizes the assignment of nine pre-specified data partitions among the three process partitions for 
the $\alpha$-shape parameter.  
The number in the upper left of the panel indicates the number of partition schemes in the $95\%$ credible set ({\it i.e.}, the 95\% CI) for this parameter.

\begin{figure}[h] 
\centering 
\includegraphics[angle=90, height=175mm]{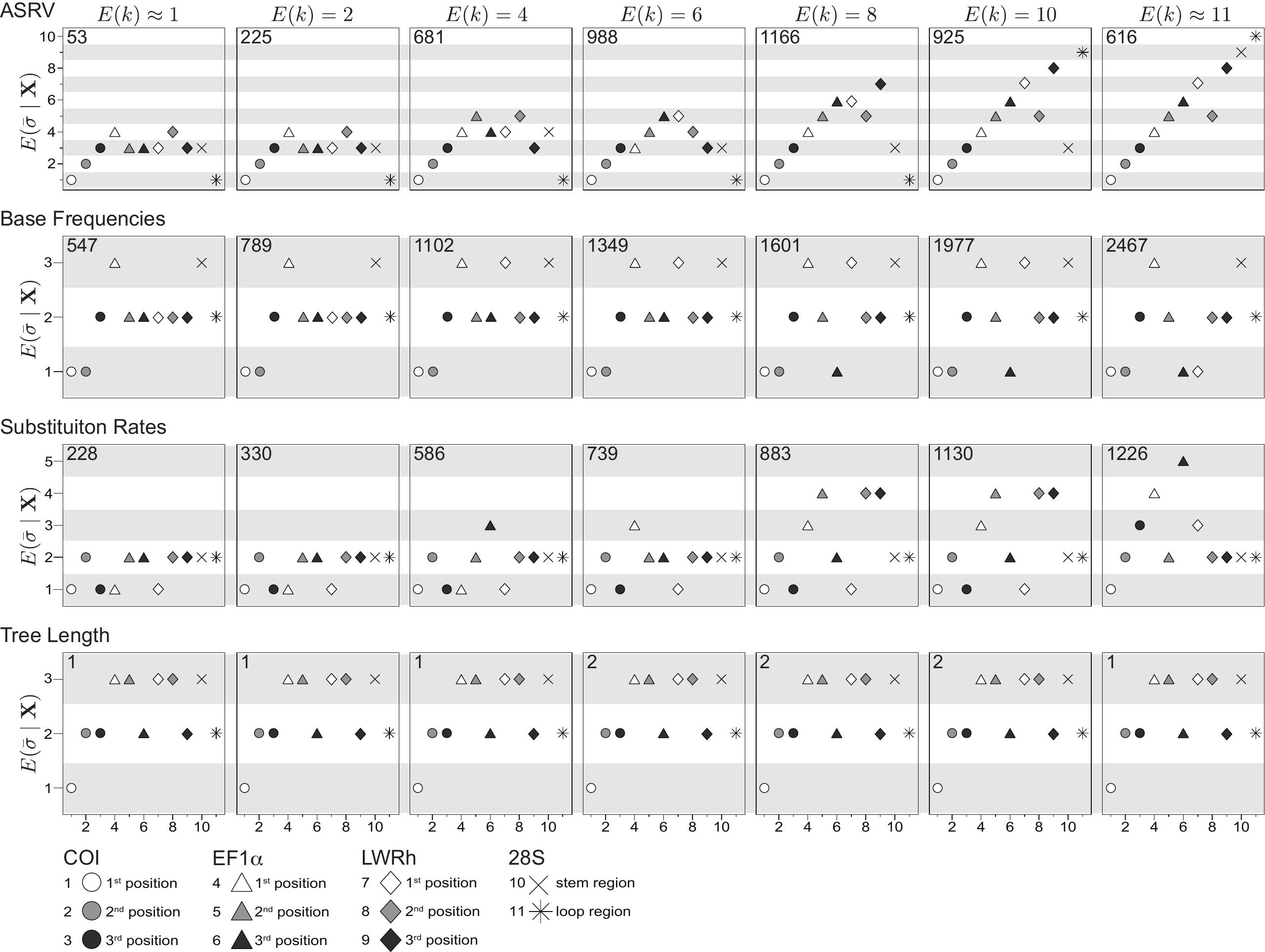} 
\caption{Inferred patterns of process heterogeneity within the gall-wasp sequence alignment inferred under the Dirichlet process prior model.  Each row of panels corresponds to a parameter of the model, and each column of panels corresponds to a specified value for the prior mean on number of process partitions in the sequence alignment, $E(k)$.  Accordingly, each panel summarizes estimates for a particular parameter under a specified value of $E(k)$.  The number of occupied rows within a panel corresponds to the estimated number of process partitions, $E(k \mid \mathbf{X})$, and the allocation of data partitions (depicted as symbols) among the rows of a panel specifies the inferred mean partition scheme, $E(\bar{\sigma} \mid \mathbf{X})$.  The number in the upper left of each panel specifies the size of the $95\%$ credible set of partition schemes.}
\label{gall_wasps}
\end{figure}

\begin{figure}[h] 
\centering 
\includegraphics[angle=90, height=175mm]{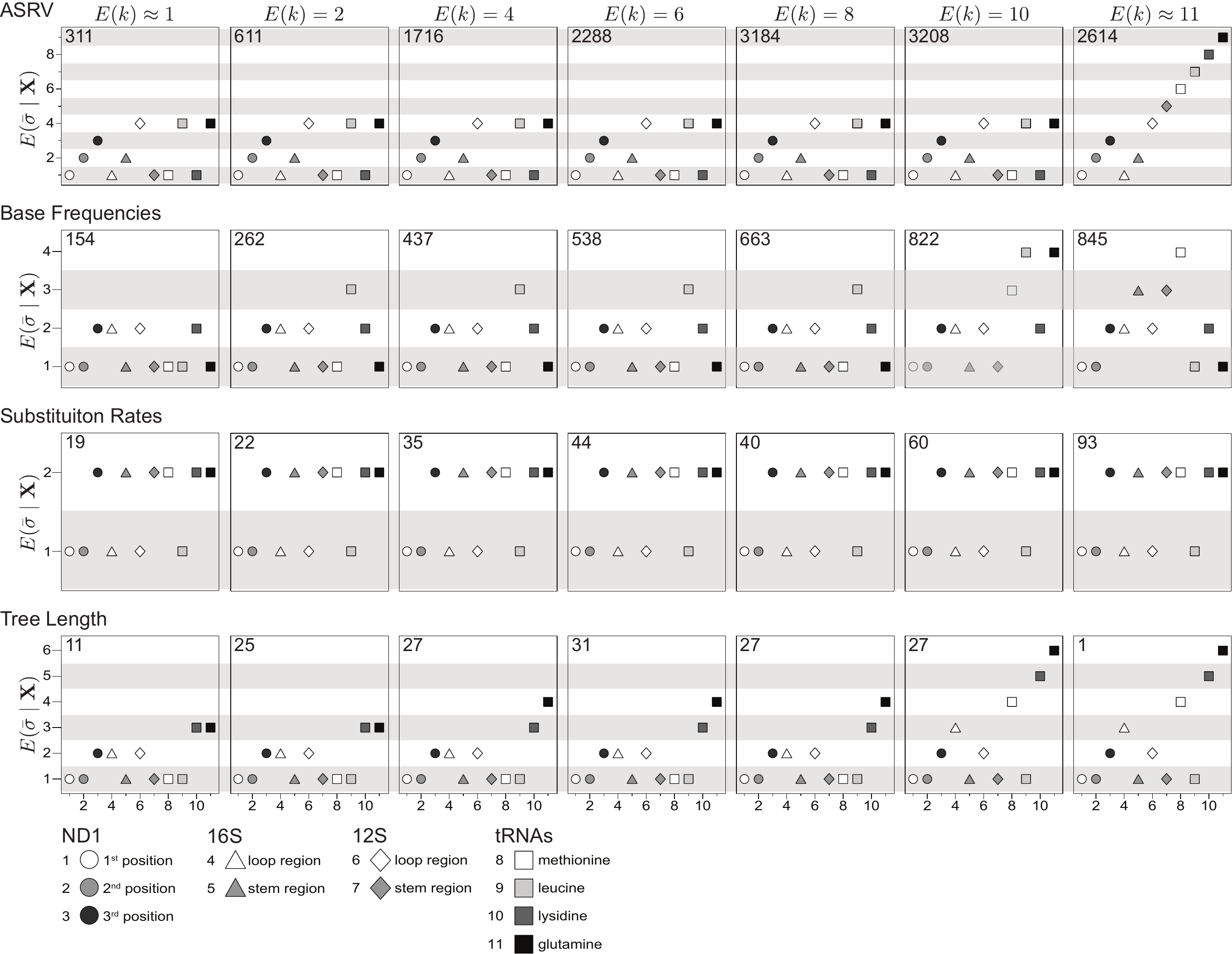} 
\caption{Inferred patterns of process heterogeneity within the skink sequence alignment inferred under the Dirichlet process prior model.  Graphical conventions as detailed in Figure \ref{gall_wasps}.}
\label{skinks}
\end{figure}

\begin{figure}[h] 
\centering 
\includegraphics[angle=90, height=150mm]{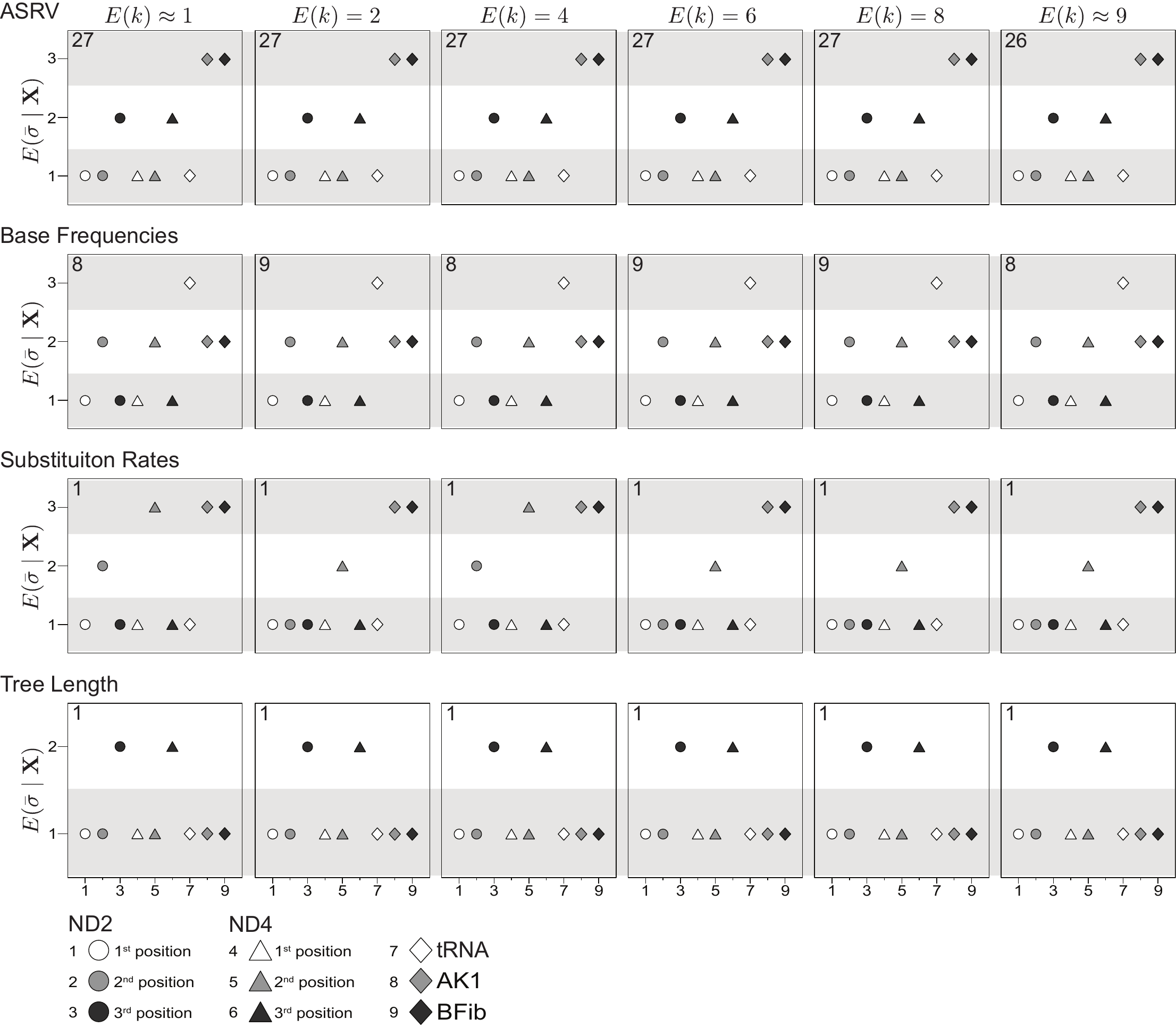} 
\caption{Inferred patterns of process heterogeneity within the hummingbird sequence alignment inferred under the Dirichlet process prior model.  Graphical conventions as detailed in Figure \ref{gall_wasps}.}
\label{hummers}
\end{figure}

\bigskip
\noindent
{\it General aspects of the results.}---Inspection of the number of process partitions comprising the mean partition schemes provides insight into the level and nature of process heterogeneity within the sequence alignments.  
For example, the $\alpha$-shape parameter was inferred to have the largest number of process partitions across all three datasets (Figures \ref{gall_wasps}--\ref{hummers}), which is consistent with previous empirical \citep[{\it e.g.},][]{nylander04} and theoretical \citep[{\it e.g.},][]{yang96a,huelsenbeck04b} results that have identified among-site substitution rate variation 
as a critically important modeling component.  
By contrast, the rank order in the number of process partitions---and the corresponding level of process heterogeneity---inferred for the other parameters (base frequencies, 
substitution rate, and tree length) differed among the three sequence alignments. 

Several generalities regarding the impact of the concentration parameter merit comment.  
First, even when a substantial proportion ($>95\%$) of the prior probability was placed on a homogeneous substitution model [{\it i.e.}, where $E(k) \approx 1$; the left-most columns in Figures \ref{gall_wasps}--\ref{hummers}], the mean partition schemes for all datasets were nevertheless inferred to have $E(k \given \mathbf{X}) > 1$, suggesting that these alignments are very difficult to explain using a uniform model.

Second, inferences of the mean partition schemes for the three datasets differed somewhat in their sensitivity to the prior on the number of partitions, $E(k)$.  
Specifically, the mean partition scheme inferred for skinks ($K = 11$) and hummingbirds ($K = 9$) were largely stable over the entire range of priors 
placed on the number of process partitions (Figures \ref{skinks} and \ref{hummers}, respectively), whereas those for gall wasps ($K = 11$) exhibited greater sensitivity to the prior on $E(k)$, such that the number of process partitions increased as more prior probability was placed on more parameter-rich partition schemes (Figure \ref{gall_wasps}).  
Even in these cases, however, sensitivity of the mean partition schemes to the priors on $E(k)$ for gall wasp dataset was restricted to two parameters, while the inferred process partitions for the other parameters remained stable over the entire range of $E(k)$.

Third, the pattern and level of uncertainty in estimates of the mean partition schemes---reflected in the size of the 95\% CI of $E(\bar{\sigma} \given \mathbf{X})$---varied somewhat with the prior mean on the number of process partitions across individual parameters and among the three datasets.  
A prominent pattern (exhibited by $40\%$ of the parameters) involved a uniform level of uncertainty in the mean partition scheme over the range of priors on $E(k)$; {\it e.g.}, the size of the 95\% CI for the mean partition scheme of the tree-length parameter in the gall wasp alignment included $1-2$ partition schemes over $E(k) = \{1.0-11.0\}$ (Figure \ref{gall_wasps}).  
An equally common pattern (also exhibited by $40\%$ of the parameters) involved an increasing level of uncertainty in the mean partition scheme that scaled linearly with the prior mean on the number of partition schemes; {\it e.g.}, the $95\%$ CI of the substitution-rate parameters in the gall wasp alignment peaks at $E(k) \approx 11$ (Figure \ref{gall_wasps}).  
As $E(k)$ increases, more prior mass is placed on partition schemes that include a greater number of process partitions, increasing the probability that data partitions will be distributed over a greater number of process partitions, such that fewer data are available to estimate each parameter.  
Accordingly, the marginal posterior probability densities for each parameter are apt to become more diffuse (especially for smaller data partitions and/or those with limited variation), which impedes the ability of the Dirichlet process prior to unambiguously assign these marginal posterior probability densities to process partitions, with a corresponding increase in the uncertainty associated with the inferred mean partition schemes. 
In the final pattern (exhibited by the remaining $20\%$ of the parameters), uncertainty in the mean partition scheme was greatest at intermediate values of $E(k)$; {\it e.g.}, the $95\%$ CI for the $\alpha$-shape parameter in the gall wasp alignment peaks at $E(k) = 8$ (Figure \ref{gall_wasps}).  
This pattern may partly reflect the increased number of possible partition schemes at intermediate values of $E(k)$.  
For example, the $11$ data partitions in the gall wasp alignment can be combined in $1,023$ unique partition schemes with two process partitions, $179,487$ unique partition schemes with six process partitions, and $11,880$ unique partition schemes with eight process partitions.  

Finally, posterior estimates of the mean partition scheme for the empirical datasets were associated with considerable uncertainty: on average, the $95\%$ CI of partition schemes included $\approx$1802 process partitions ({\it i.e.}, calculated by summing the size of the credible sets for the four parameters---the columns in Figures \ref{gall_wasps}--\ref{hummers}---averaged over all values of $E(k)$ and over the three alignments).  
The generally large number of plausible partition schemes for the empirical datasets contrasts sharply with those for the simulated data. 
The large size of the empirical credible sets for partition schemes indicates a correspondingly high degree of uncertainty in the choice of mixed model, and emphasizes the appeal of integrating inferences of phylogeny (and other model parameters) over this important source of uncertainty (rather than conditioning inferences on any particular partition scheme).

\newpage

\begin{center}
{\sc Discussion}
\end{center}

\bigskip
\noindent
{\it Conceptual relationship of the DPP model to other approaches.}---Accommodating variation in the substitution process across an alignment of nucleotide sequences is inherently a partitioning problem:
the objective is to identify subsets of sites that share common aspects of their molecular evolution.
This requires specifying or estimating both the number of process partitions, $k$, and the allocation vector, ${\mathbf z}$, specifying how the data elements are assigned to those $k$ process partitions.
This problem has been addressed using three distinct approaches.
The \emph{mixed-model} approach \citep[{\it e.g.},][]{ronquist03} specifies both the number of process partitions and the assignment of data elements to those process partitions, which collectively defines a partition scheme.
Alternative partition schemes represent competing explanations (mixed models) of the pattern of process heterogeneity that gave rise to the data.
The average fit of each candidate mixed model to the data is first assessed by estimating its marginal likelihood.
The marginal likelihoods of the candidate models are then compared using Bayes factors to select mixed model that provides the best description of the process heterogeneity in the dataset.

By contrast, \emph{finite-mixture} models specify only the number of process partitions, $k$, but treat the allocation vector, ${\mathbf z}$,  as a random variable.
That is, the number of process partitions is assumed, but the assignment of the $K$ data partitions to those $k$ process partitions is assumed to be unknown. 
Accordingly, the likelihood function takes the form of a weighted average over the $k$ process partitions.
Specifically, the likelihood for each of the $K$ data partitions is integrated over the $k$ process partitions, and the resulting set of $k$ likelihoods are then combined as a weighted average, where the weights are estimated from the data.
\citet{pagel04} proposed a finite-mixture model that allows for $k$ unique process partitions that differ in the exchangeability rates, while assuming that other aspects of the substitution model---the stationary frequencies, tree length, alpha-shape parameter---do not vary over the sequence alignment.
This innovative approach obviates the need to specify the allocation vector of the partition scheme, but requires that we specify the number of process partitions, $k$.
A brute force solution for specifying the number of process partitions would entail performing a series of analyses that increment the number of process partitions, $k=\{1\dots K\}$, and estimating the marginal likelihood for each of the corresponding $K$ finite-mixture models.
Bayes factors could then be used to identify the finite-mixture model that provides the best description of the process heterogeneity in the dataset.   

The Dirichlet process is an example of an \emph{infinite-mixture} model.
Under this approach, each of the $K$ data partitions is modeled as having evolved under one of a countably infinite number of processes---as having been generated by a mixture model---without knowing the form of the mixture.
Instead, the form of the mixture (including both the number of process partitions and the allocation of data elements among them) is governed by higher-level parameters  ({\it i.e.}, the concentration parameter, $\chi$, and the base distribution, $G_0$, of the DPP model).
In principle, the likelihood function involves an integral over all possible mixture models conditional on these higher-level parameters.
In practice, however, numerical methods (MCMC algorithms) are used to perform the integration.
This approach provides a dynamic solution to mixture modeling, allowing the dimensions of the model to efficiently expand or shrink as dictated by patterns in the data.

\bigskip
\noindent
 {\it Practical comparison of the DPP and mixed-model approaches.}---The Dirichlet process prior approach for accommodating process heterogeneity is most similar in spirit to the conventional partition-scheme selection that relies on Bayes factors.  
The three alignments evaluated in the present study have previously been subjected to extensive mixed-model selection using Bayes factors, which affords the opportunity to compare these two approaches.  
Below we consider four aspects of this comparison.

First, in every case, analyses using the Dirichlet process prior model discovered novel mean partition schemes that were not considered in the previous studies that evaluated candidate partition schemes using Bayes factors.  
We emphasize that the selected studies are notable for the relatively large number of partition schemes that they considered.  
Nevertheless, the need to perform a full MCMC analysis under each candidate partition scheme (in order to estimate the corresponding marginal likelihood) entails substantial computational overhead.
Accordingly, even these exceptionally thorough studies necessarily considered only a small fraction of the possible mixed-model space.  
Specifically, the hummingbird study evaluated 0.0000000000000045\% (9 of $2$x$10^{17}$) of the possible process partition schemes for an alignment with $K=9$ data partitions, and the gall wasp and skink studies evaluated 0.0000000000000000000042\% ($9$ of $2$x$10^{23}$) of the partition schemes possible for an alignment with $K=11$ data partitions.
The ability of the Dirichlet process prior model to integrate inference of phylogeny over the entire range of partition schemes in the course of a single MCMC analysis greatly increases the probability of estimating under the partition scheme that best captures process heterogeneity within the data, thereby minimizing biases associated with mixed-model misspecification.

Second, the complexity of the partition schemes preferred by the two approaches differs markedly.  
As in all other applications, mixed-model selection using Bayes factors identified the most parameter-rich candidate model for the gall wasp, skink, and hummingbird alignments.
By contrast, the mean partition schemes inferred under the Dirichlet process prior approach were considerably more parsimonious.  
Specifically, the mean/selected partition schemes identified under the Dirichlet process prior/Bayes factor approaches included 32/99 free parameters for the gall wasp alignment, 27/98 free parameters for the skink alignment, and 29/108 free parameters for the hummingbird alignment.  
The strong preference for the most complex candidate mixed model again illustrates the well-known bias of Bayes factor comparisons based on marginal likelihoods inferred using the harmonic-mean estimator \citep[{\it e.g.},][]{lartillot06, brown07, fan11, bael12, bael13}. 
More reliable (but substantially more computationally intensive) marginal-likelihood estimators are available \citep[{\it e.g.},][]{lartillot06, xie11, fan11, bael12, bael13, ronquist12}.
However, the scope of the problem may render these solutions impractical.
For example, even if we could estimate the marginal likelihoods at the (computationally astronomical) rate of one partition scheme per second, it would still take $\approx 6,723,137,822,834,178$ years to evaluate all possible partition schemes for the gall wasp alignment (roughly one million times the age of the universe). 

A more fundamental limitation of the conventional approach for selecting partition schemes pertains to (mixed) model uncertainty.
Even if we could feasibly evaluate all possible candidate partition schemes for a given alignment and reliably select the mixed model that provides the best fit to that dataset, it may nevertheless be imprudent to condition inference on the chosen partition scheme.
For example, the $95\%$ credible set of partition schemes for the datasets evaluated here comprised $2935$, $2434$, and $37$ mixed-models for the gall wasp, skink, and hummingbird analyses, respectively.
Selecting {\it any single partition scheme}---even the MAP estimate---would ignore the considerable uncertainty regarding the choice of mixed model that gave rise to these alignments.
Failure to accommodate uncertainty in the choice of (mixed) model is likely to bias our phylogenetic estimates \citep[{\it cf.},][]{huelsenbeck04d, li12}.
Accordingly, our phylogenetic estimates will be rendered more robust if we accommodate this mixed-model uncertainty by integrating over  all possible partition schemes using an approach such as the Dirichlet process  model.
Although Bayes factors may not provide a viable approach for \emph{accommodating} process heterogeneity, we nevertheless envisage them playing a key role in testing specific hypotheses regarding process heterogeneity, {\it e.g.}, comparing two competing hypotheses regarding the assignment of data elements to process partitions.

The final point of comparison between the DPP and mixed-model approaches pertains to difference in error variance.  
A corollary of the substantial reduction in model complexity under the Dirichlet process prior model---achieved by reduction of superfluous parameters---is a corresponding reduction in the error variance and associated uncertainty of parameter estimates.  
For example, the level of uncertainty associated with the topology parameter---as reflected in the size of the 95\% credible set of trees---estimated under the Dirichlet process prior/Bayes factor approaches included $927/1513$ trees for the gall wasp alignment, $2488/3106$ trees for the skink alignment, and $3961/4672$ trees for the hummingbird alignment, respectively. 
The reduction in estimation error (particularly the topology and branch-length parameters) under the Dirichlet process prior model should decrease uncertainty not only in estimates of phylogeny, but also in phylogeny-based inferences focused on the study of character evolution, rates of lineage diversification, biogeographic history, etc.
A model that more realistically describes the processes by which increasingly complex, genome-scale sequence data arise will provide more reliable estimates of phylogeny and, by extension, more reliable inferences of evolutionary processes from phylogenies. 

\bigskip
\noindent
{\it Applications and extensions.}---The Dirichlet process prior model we describe provides a versatile approach for accommodating process heterogeneity across an alignment of nucleotide sequences, and so should both improve our estimates of phylogeny and inferences based on those phylogenies. 
Importantly, this framework can readily be applied to other inference problems and extended in various ways.
For example, the DPP approach could be productively applied to numerous problems in molecular evolution simply by reversing the relationship between focal and nuisance parameters.
That is, our motivation for the DPP approach was to provide more robust estimates of phylogeny (tree topology and branch lengths) by virtue of integrating over all possible patterns of process heterogeneity across the alignment (where the process partitions are effectively nuisance parameters).
However, an equally fruitful application of the DPP approach would focus directly on the nature of process heterogeneity to study various aspects of molecular evolution (where the phylogeny is essentially a nuisance parameter), providing a flexible framework to directly study how various aspects of the substitution process---stationary frequencies, exchangeability rates, overall substitution rate and the degree of ASRV---vary within and among gene/omic regions (\emph{c.f.}, Figure \ref{results_example}).

The DPP approach can also be extended in various ways by relaxing some current assumptions.
For example, we have implemented the DPP approach for accommodating process heterogeneity under the so-called `unconstrained' phylogenetic model \citep[{\it e.g.},][]{Heath14}, where the rate of substitution is independent on each branch, and the branch length is proportional to the degree of molecular evolution.
Because accommodating process heterogeneity is known to be important for estimating divergence times \citep[{\it e.g.},][]{marshall06,poux08,vendetti08}, it may also be useful to implement the DPP approach under a `constrained' phylogenetic model (where the rate of substitution is shared across branches according to a strict-/relaxed-molecular clock model) and branch lengths are proportional to relative or absolute time. 
In lieu of a fully hierarchical solution for estimating divergence times, the current method could be used in a serial Bayesian approach; {\it i.e.}, where \verb!AutoParts! is first used to estimate the mean partition scheme that best describes the nature of process heterogeneity across the alignment, and then this mixed substitution model is used to estimate divergence times (under the any strict-/relaxed clock models). 

Our approach accommodates variation in the evolutionary process across a sequence alignment, but nevertheless assumes that these sequences share a common phylogenetic history. 
For some datasets---particularly multiple nuclear loci sampled from closely related and/or recently diverged species---the assumption of a single phylogenetic history is likely to be violated by processes such lineage sorting and deep coalescence.  
Accordingly, it may prove valuable to extend the Dirichlet process prior model to accommodate this aspect of process heterogeneity by treating the tree topology like the other parameters of the DPP model \citep[{\it c.f.}, ][]{ane07}.  
That is, rather than assuming that all data partitions share a common tree-topology parameter, we could treat the number of topologies and the assignment of data partitions to those tree topologies as random variables.  
We are optimistic that the Dirichlet process prior approach as currently described will provide biologists with an efficient and robust method for accommodating process heterogeneity when estimating phylogeny, and that future extensions of this framework will prove fruitful for addressing related problems in phylogenetic biology.

\bigskip

\begin{center}
{\sc Acknowledgments}
\end{center}

\noindent
We are grateful to members of the Moore and Huelsenbeck labs---particularly, Mike May, Tracy Heath, Sebastian H\"ohna, Michael Landis, and Bastien Boussau---for thoughtful discussion of this work.
We also wish to thank Nicolas Rodrigue for offering an exceptionally helpful review, and for encouraging us to discuss the relationship between the DPP approach and finite mixture models.
This research was supported by grants from the NSF (DEB-0445453) and NIH (GM-069801) awarded to J.P.H., and by NSF grants (DEB-0842181, DEB-0919529, and DBI-1356737) awarded to BRM.
Computational resources for this work were provided by an NSF XSEDE grant (DEB-120031) to BRM.

\newpage
\bibliography{auto_parts}

\newpage
\begin{table}[h!]
\caption{Performance (coverage\textsuperscript{1} [precision\textsuperscript{2}]) of the DPP approach under a typical simulation.}
\centering
\begin{tabular}{lllllll}
\toprule
& \multicolumn{3}{c}{$\chi$ (fixed)\textsuperscript{3}}  & \multicolumn{3}{c}{$\chi$ (gamma)\textsuperscript{3}} \\
\cmidrule(r){2-4} \cmidrule(r){5-7}
Parameter & $E(k) \approx 1$ & $E(k) =3$ & $E(k) \approx 6$ & $E(k) \approx 1$ & $E(k) =3$ & $E(k) \approx 6$ \\ 
\midrule
ASRV & 1.0 [99.8] & 1.0 [98.9] & 1.0 [97.7] & 1.0 [99.8] & 1.0 [98.9] & 1.0 [97.7] \\ 
\rowcolor{gray!20}
Base frequencies & 1.0 [99.8] & 1.0 [98.9] & 1.0 [97.7] & 1.0 [99.8] & 1.0 [98.9] & 1.0 [97.7] \\ 
Substitution rates & 1.0 [100.0] & 1.0 [100.0] & 1.0 [100.0] & 1.0 [100.0] & 1.0 [100.0] & 1.0 [100.0]  \\ 
\rowcolor{gray!20}
Tree length & 1.0 [98.3] & 1.0 [97.8] & 1.0 [98.2] & 1.0 [98.3] & 1.0 [97.8] & 1.0 [98.2]  \\ 
\end{tabular}

\medskip
{\fontsize{9}{9}\selectfont
\begin{adjustwidth}{.2cm}{.1cm}
These results correspond to the simulation experiment described in Figure \ref{sim_design}.\\
\textsuperscript{1}The probability that the $95\%$ credible interval (CI) includes the true (generating) partition scheme.\\
\textsuperscript{2}The mean number of partition schemes comprising the $95\%$ CI as a function of the maximum possible size.\\
\textsuperscript{3}The concentration parameter, $\chi$, specifying the prior mean on the number of partition schemes, $E(k)$, is either fixed or governed by a gamma hyperprior.
\end{adjustwidth}}
\label{tab:simulation}
\end{table}

\newpage

\begin{center}
{\sc Figure Legends}
\end{center}

{\fontsize{10}{10}\selectfont
\noindent {Figure \ref{sim_design}}. A typical simulation experiment.  This simulation involves $K = 6$ data partitions, each with $L = 1000$ sites. All data partitions are assumed to share a common tree topology, $\tau$, and branch-length proportions, ${\mathbf p}$, but other aspects of the substitution process may differ between data partitions, including: the degree of among-site rate variation (ASRV, governed by the $\alpha$-shape parameter); the vector of nucleotide frequencies, \mbox{\boldmath$\pi$\unboldmath}; the vector of exchangeability rates, ${\mathbf r}$, and the overall rate of substitution/tree length, $T$.  The partition scheme, $\sigma$, for each of the four parameters specifies both the number of process partitions, $k$, and the allocation vector, ${\mathbf z}$, of the $K = 6$ data partitions to those $k$ process partitions. For example, the 6 data partitions are allocated to $k = 3$ distinct process partitions for the $\alpha$-shape parameter (the shaded rows of the panel), where the allocation vector is ${\mathbf z}  = (1,2,3,3,2,1)$ (depicted by the assignment of the data-partition symbols to their respective rows), and the specific parameter values for each process partition are indicated in the corresponding rows of the panel.

\bigskip

\noindent {Figure \ref{results_example}}. Inferred patterns of among-site substitution rate variation within the hummingbird sequence alignment.  (A) Marginal posterior probability densities for the $\alpha$-shape parameter of the discrete gamma model have been estimated for each of the 9 predefined data partitions.  (B) In contrast to the unimodal marginal densities for other data partitions, the posterior density for the second-position sites of the ND4 gene is distinctly bimodal, suggesting residual process heterogeneity within this data partition.  (C) Analyses under the Dirichlet process prior model provide estimates both for the number of process partitions, $E(k \mid \mathbf{X}) = 3$ (represented as rows in the panel), as well as the mean partition scheme, $E(\bar{\sigma} \mid \mathbf{X})$ (depicted by the assignment of the data-partition symbols to their respective rows).

\bigskip

\noindent {Figure \ref{gall_wasps}}. Inferred patterns of process heterogeneity within the gall-wasp sequence alignment inferred under the Dirichlet process prior model.  Each row of panels corresponds to a parameter of the model, and each column of panels corresponds to a specified value for the prior mean on number of process partitions in the sequence alignment, $E(k)$.  Accordingly, each panel summarizes estimates for a particular parameter under a specified value of $E(k)$.  The number of occupied rows within a panel corresponds to the estimated number of process partitions, $E(k \mid \mathbf{X})$, and the allocation of data partitions (depicted as symbols) among the rows of a panel specifies the inferred mean partition scheme, $E(\bar{\sigma} \mid \mathbf{X})$.  The number in the upper left of each panel specifies the size of the $95\%$ credible set of partition schemes.  

\bigskip

\noindent {Figure \ref{skinks}}. Inferred patterns of process heterogeneity within the skink sequence alignment inferred under the Dirichlet process prior model.  Graphical conventions as detailed in Figure \ref{gall_wasps}.

\bigskip

\noindent {Figure \ref{hummers}}. Inferred patterns of process heterogeneity within the hummingbird sequence alignment inferred under the Dirichlet process prior model.  Graphical conventions as detailed in Figure \ref{gall_wasps}.

}

\newpage

\newpage

\newpage

\newpage

\newpage

\end{document}